\documentclass[%
 reprint,
superscriptaddress,
 amsmath,amssymb,
 aps,
]{revtex4-1}

\usepackage{graphicx}% Include figure files
\usepackage{dcolumn}% Align table columns on decimal point
\usepackage{bm}% bold math
\usepackage{mathptmx}
\usepackage{textcomp}

\begin{document}

\preprint{AIP/123-QED}

\title{Electric-field-driven conductance switching in encapsulated graphene nanogaps %\emph{fabricated with hydrogen silsesquioxane?}%
}
% Force line breaks with \\

\author{Eugenia Pyurbeeva}
\email{e.d.pyurbeeva@qmul.ac.uk}
\affiliation{ 
School of Physics and Astronomy, Queen Mary University of London, Mile End Road, London, E1 4NS, UK}
\author{Jacob L. Swett}
\affiliation{Department of Materials, University of Oxford, Oxford, OX1 3PH, UK}
\author{Qingyu Ye}
\affiliation{School of Biological and Chemical Sciences, and Materials Research Institute
Queen Mary University of London ,Mile End Road, London E1 4NS, UK}
\author{Oscar W. Kennedy}
\affiliation{London Centre for Nanotechnology and Department of Physics and Astronomy, University College London, London, United Kingdom}
\author{Jan A. Mol}
\email{j.mol@qmul.ac.uk}
\affiliation{ 
School of Physics and Astronomy, Queen Mary University of London, Mile End Road, London, E1 4NS, UK}

\date{\today}% It is always \today, today,

\begin{abstract}
Feedback-controlled electric breakdown of graphene in air or vacuum is a well-established way of fabricating tunnel junctions, nanogaps, and quantum dots. We show that the method is equally applicable to encapsulated graphene constrictions fabricated using hydrogen silsesquioxane. The silica-like layer left by hydrogen silsesquioxane resist after electron-beam exposure remains intact after electric breakdown of the graphene. We explore the conductance switching behavior that is common in graphene nanostructures fabricated via feedback-controlled breakdown, and show that it can be attributed to atomic-scale fluctuations of graphene below the encapsulating layer. Our findings open up new ways of fabricating encapsulated room-temperature single-electron nanodevices and shed light on the underlying physical mechanism of conductance switching in these graphene nanodevices.
\end{abstract}

\maketitle

Electric breakdown of graphene \cite{Prins2011, Lau2014, Sadeghi2015} provides an inexpensive and facile way of creating two-dimensional nanostructures, including tunnel junctions \cite{Prins2011, Lau2014, Sadeghi2015, Limburg2018} and quantum dots \cite{Harzheim2020}. While these structures have mainly been studied in the context of electrodes for single-molecule devices \cite{Prins2011}, they also hold potential as active electronic elements in their own right as single electron transistors \cite{Barreiro2012, Puczkarski2015, Harzheim2020} and for phase-change memory \cite{Posa2017}. A key challenge in harnessing this potential is the requirement to integrate electric breakdown with additional pre- and post-processing fabrication steps \cite{Han2014}.

As graphene electronic devices approach practical applicability, encapsulation can be beneficial, reducing the probability of damaging the graphene in subsequent fabrication steps \cite{Li2019}. As 2D materials have a large surface to volume ratio, surface contamination in the active region of the device can result in large variations of conductance \cite{Ray2016}. While environment-dependent conductance has succesfully been utilised to detect gas molecules \cite{Ray2016, Rani2019} and DNA \cite{Mishra2020}, it should be avoided in most applications to minimise device-to-device variability.

Patterning graphene on a silica substrate with hydrogen silsesquioxane (HSQ) enables direct integration of a silica-like encapsulation layer in the electron beam lithography process, creating sandwiched graphene nanostructures. We demonstrate that the electric breakdown process to fabricate these nanostructures is equally efficient for sandwiched and open graphene devices.

To benchmark the stability of encapsulated graphene devices, we compare the observed conductance switching to that previously reported in open devices \cite{Sarwat2017, Posa2017}. We attribute the electric-field-induced switching to atomic-scale fluctuations within the silica-like encapsulation. While this switching behavior is likely to limit the use of encapsulated graphene nanostructures for some electronic applications, its understanding can offer insights into the miniaturisation limits of graphene electronics, while additionally, it opens a viable route towards the development of nanoscale non-volatile memory elements \cite{Posa2017, Fried2020, Posa2021}.

HSQ has been previously used for the fabrication of graphene devices \cite{Han2007, Chen2007}, however the developed resist is typically removed after patterning with hydrofluoric acid for contact deposition or further doping \cite{Chen2007}. Recent work has also shown quantum dot formation through effective doping of graphene with varying exposure dose \cite{Wang2019}. It has been demonstrated that HSQ encapsulated devices have an increased conductance and yield compared to those fabricated with polymethyl methacrylate (PMMA) resist due to the graphene being protected during post-exposure processing and the decreased contamination and doping of graphene by resist residue \cite{Li2019}.

%%%%%%%%%%%%%%%%%%%%%%%%%%
\begin{figure}
\includegraphics[width=0.5\textwidth]{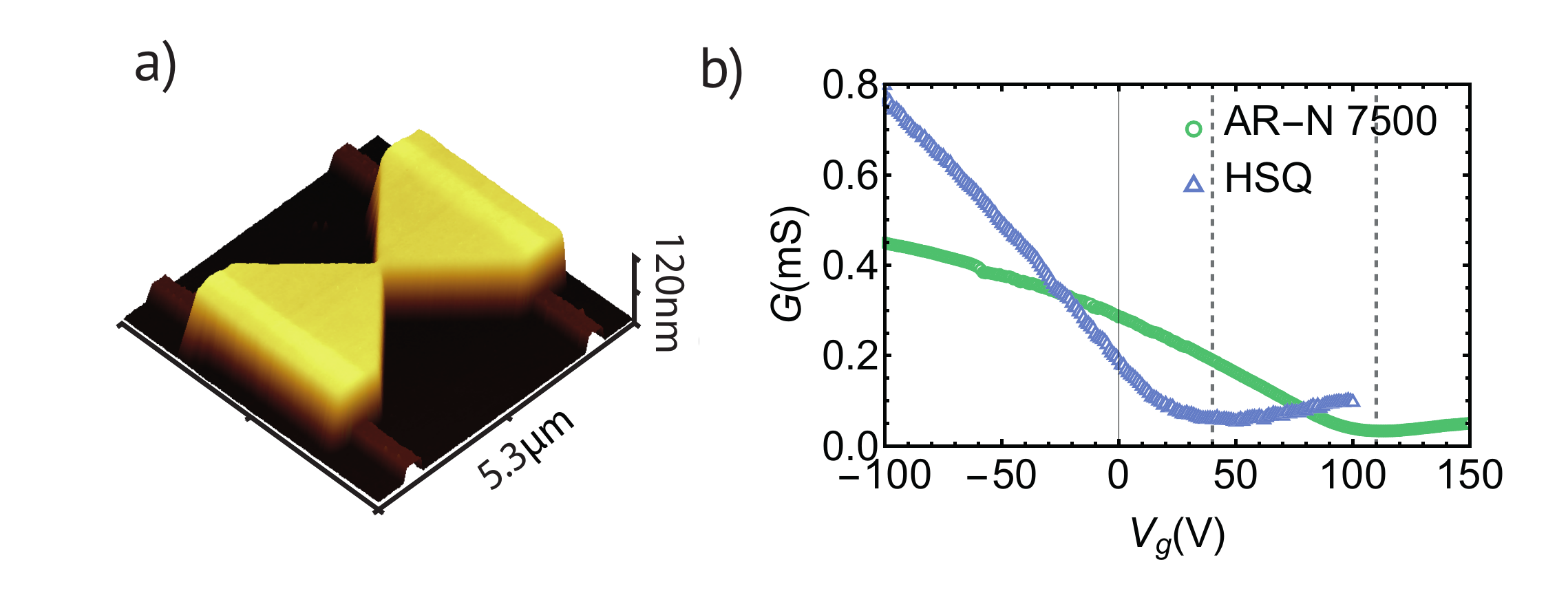}\caption{\label{fabrication} a) 3D rendered AFM image of the fabricated device. A bowtie-shaped graphene constriction overlaps two pre-fabricated gold electrodes. The thickness of the residual HSQ layer is approximately 100nm.  b) Conductance dependencies on gate voltage of the non-electroburnt constrictions fabricated with AR-N 7500 (graphene is uncovered) vs constriction fabricated with HSQ. The latter shows a significantly lower shift of the Dirac point (conductance minimum) from zero voltage.}
\end{figure}
%%%%%%%%%%%%%%%%%%%%%%%%%%

We fabricate devices on heavily doped (0.005 Ohm-cm ) silicon substrates with a 275 nm dry thermal oxide layer, on which electrodes have been deposited using electron-beam lithography: first 5nm Cr + 15nm Au with electron beam evaporation, followed by a photolithography step, using contact lithography and 5nm Cr + 95nm Au deposited with electron beam evaporation. The wafer was then completely covered with chemical vapour deposition grown graphene (Graphenea). Using electron beam lithography, we patterned bowtie-shaped graphene constrictions overlapping both electrodes with the width of in the narrowest point of 300nm.

The AFM image of the resulting device is shown on figure \ref{fabrication}a, where an approximately 100 nm thick bowtie-shaped silica layer can be seen overlapping with the two gold electrodes. The thickness of the residual silica-like layer covering the graphene constriction is approximately 100 nm. 

We measure the dependence of electrical conductance on the back-gate voltage, applied to the doped silicon substrate, for a graphene constriction fabricated with HSQ and compare it to the conductance of a constriction of the same geometry fabricated with an organic negative resist (AR-N 7500) which is removed after development. The results in Figure \ref{fabrication}b show the device fabricated with HSQ having a higher conductance and a smaller shift of the Dirac point (corresponding to the conductance minimum) from zero gate voltage. While absolute conductance is a poor measure of graphene contamination, as it is affected by the geometry \cite{Qi2014}, the shift of the Dirac point is a more reliable quantity. The Dirac point is typically shifted away from zero gate voltage in graphene devices due to interaction with the substrate and contamination and effective doping of graphene with the resist residue \cite{Wang2019}. In the device fabricated with HSQ this shift is several times smaller than in the device fabricated with AR-N 7500 ($\sim 40$V of gate voltage compared to $\sim 120$V) which indicates that the encapsulated graphene is less contaminated. Both results agree with the previous studies \cite{Li2019, Wang2019}.

%%%%%%%%%%%%%%%%%%%%%%%%
\begin{figure}
\includegraphics[width=0.5\textwidth]{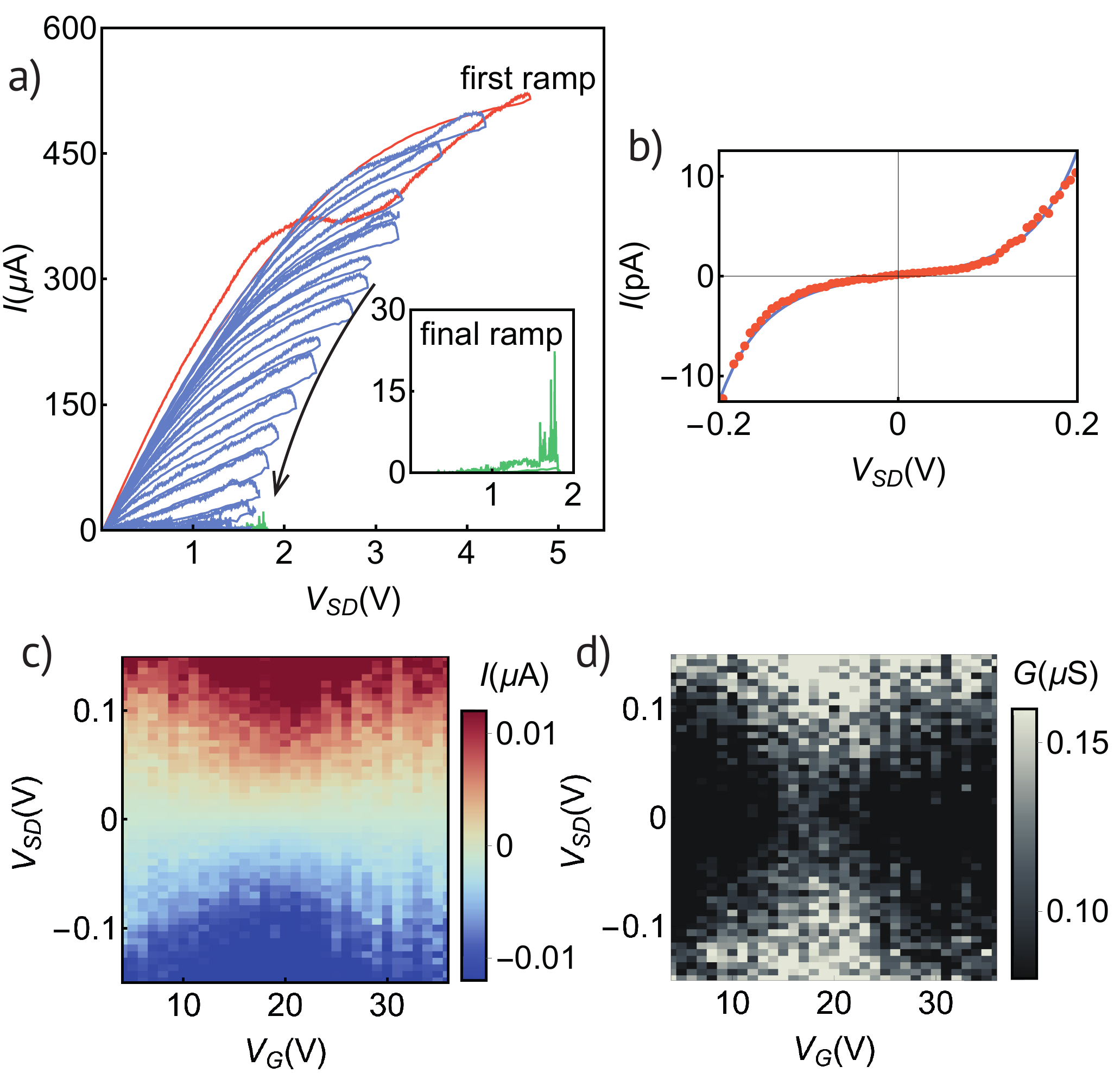}\caption{\label{electroburning} a) Current-voltage (IV) traces during the feedback-controlled electroburning process. The first trace (shown in red) shows a dip in conductance on the upwards ramp of the voltage, which is usual for electric breakdown. The following ramps result in gradually decreasing conductance -- indicated by arrow. The inset shows sudden spikes in conductance immediately before the gap formation that are attributed to due to the transition between two-path and single-path conductance. b) IV trace of an electroburnt device fitted to the Simmons model. The gap width is approximately 2nm.  c) Current stability diagram of a encapsulated device showing Coulomb blockade. d) Differential conductance stability diagram of the same device. }
\end{figure}
\begin{figure}
\includegraphics[width=0.5\textwidth]{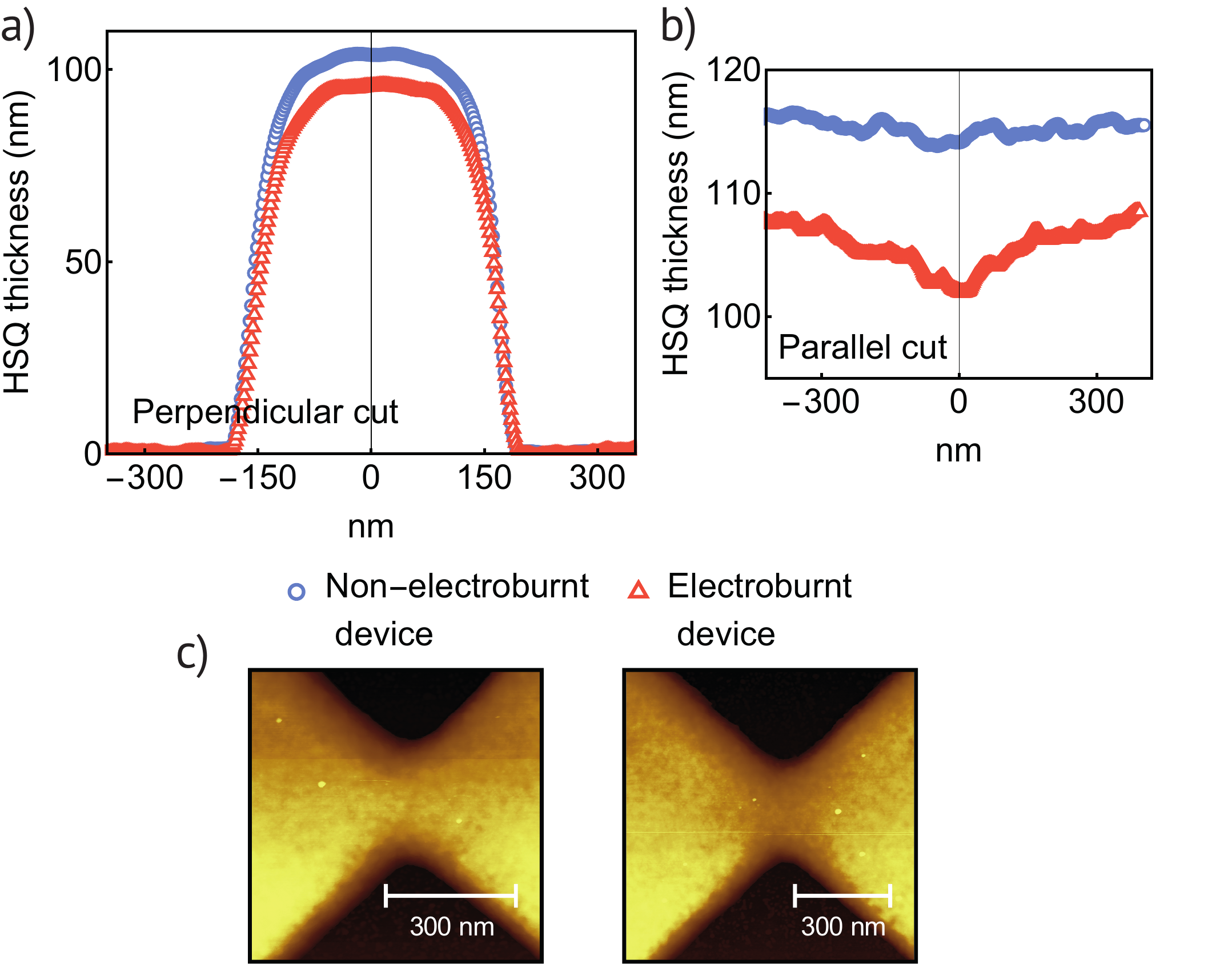}\caption{\label{afm} Compared height profiles of a device on which electric breakdown has not and has been performed taken by AFM where the origin is the the centre of the constriction. The profile cuts perpendicular to the constriction show that after electroburning the device remains well-encapsulated. In the cuts parallel to the constriction a slight dip can be seen in the device after electric breakdown, coinciding with the centre of the constriction.  Close-up AFM images of the centre of the devices are also shown.}
\end{figure}
%%%%%%%%%%%%%%%%%%%%%%%%

We perform feedback-controlled electric breakdown on the encapsulated graphene constriction by repeatedly ramping up the voltage across the constriction up to a point of a sudden drop in conductance, signifying the start of graphene breakdown \cite{Lau2014}. After the conductance drop, the voltage is quickly ramped down. Multiple voltage ramps lead to gradual narrowing of the constriction, which can result in the formation of a nanogap with a characteristic size of of several nanometres \cite{Limburg2018} or a quantum dot of similar characteristic size \cite{Barreiro2012, Harzheim2020, Fried2020}, which has been attributed to a small graphene island remaining in the gap.  However, due to the size of the system, the exact structure of the dot is not known.

The electric breakdown process for the devices fabricated with HSQ is very similar to the non-encapsulated devices. We performed feedback-controlled breakdown on both at room temperature in ambient conditions and were able to use identical parameters -- positive and negative ramp rate (5–8 V/s and 2300–5000V/s ), maximum voltage (10V) and target resistance (500 M$\Omega$). The breakdown success rate was comparable to that of non-encapsulated graphene devices  -- 64\% exhibited non-linear current-voltage characteristics indicative of nanogaps compared to 71\% in \cite{Lau2014}.

A common feature of the electric breakdown process is a significant conductance enlargement in the end of the last cycle, which is attributed to a quantum interference effect due to the transition between multi-path and single path conductivity in the remaining carbon chains bridging the gap \cite{Sadeghi2015}. This effect is also present in encapsulated devices as shown in Figure \ref{electroburning}a. In many cases, several peaks were present, which we attribute to the repeated formation and destruction of the final carbon filament. 

\begin{figure}
\includegraphics[width=0.5\textwidth]{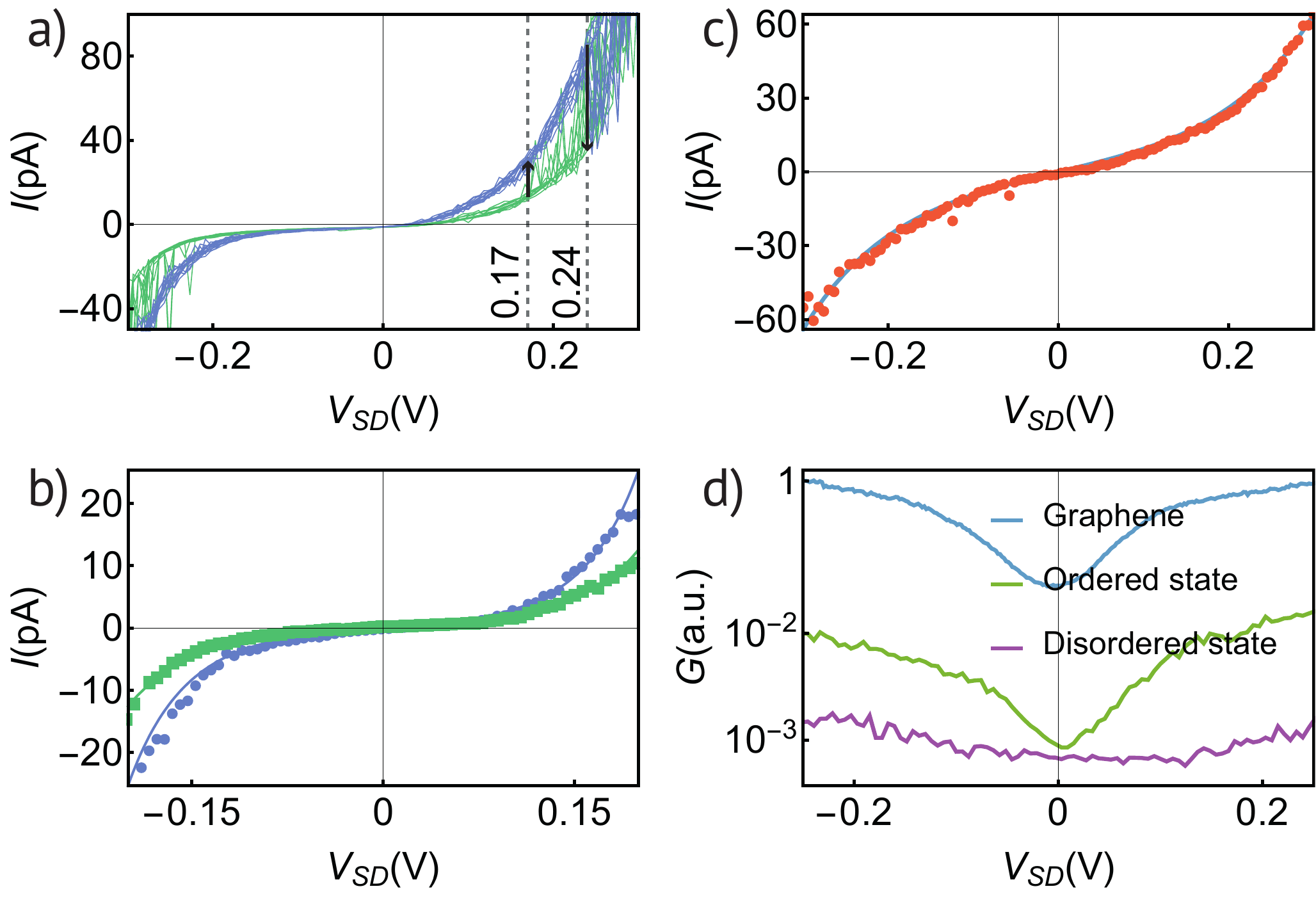}\caption{\label{switching} a) Current-voltage traces of a device switching between two conductance states. Two critical bias voltages can be seen: approximately 0.17V for a low to high switch and approximately 0.24V for the reverse switching. b) Simmons model fits to the two conductance states in figure a. The high state has a gap width several \AA narrower than the low state, indicating geometric restructuring at the atomic scale. c) IV trace of an electroburnt device before the voltage-driven restoration of ordered graphene and Simmons model fit to it, indicating a gap width of 2nm. d) Conductance-voltage traces for a graphene device before feedback-controlled electroburning (top) and two electroburnt devices after a voltage-driven switch to a high conductance state. The y-axis shows the typical conductances (in units of the typical graphene device conductance) and the traces are shifted vertically for ease of presentation.}
\end{figure}

We fit the current-voltage (IV) traces of several devices after electric breakdown to the Simmons model \cite{Simmons1963, Zimbovskaya2013, Limburg2018} (see figures \ref{electroburning}b, \ref{switching}b,c) and estimate the width of the gap to be approximately 2 nm, which agrees with the results for devices fabricated with AR-N 7500 \cite{ElAbbassi2017}.

In addition to tunnel junctions, some devices show Coulomb diamonds indicative of the formation of a quantum dot \cite{Gehring2016a} -- see figures \ref{electroburning} c,d. The estimated lower bound for the quantum dot addition energy extracted from \ref{electroburning} c,d is 200 meV, which is an order of magnitude greater than the characteristic thermal energy $k_{\rm{B}}T$for room temperature (approximately 25 meV). 

The large addition energy of quantum dots is one of the advantages of feedback-controlled electric breakdown. In order for Coulomb blockade to be observed, the energy of the electrostatic interaction of two electrons on the quantum dot has to be much greater than the characteristic thermal energy \cite{Barreiro2012}: 
\begin{equation}
	\frac{e^2}{4\pi \varepsilon_0 d}\gg k_{\rm{B}}T
\end{equation} 
where $d$ is the diameter of the quantum dot.  For Coulomb blockade to be present at room temperature, this requires $d \ll 50$ nm and therefore $d$ in the order of nanometres, which is close to the state-of-the-art in current nanofabrication techniques \cite{Vitor2016, Manfrinato2017}. As a result, the size restriction makes electroburning an attractive, cheap alternative method of fabricating quantum dots demonstrating room-temperature Coulomb blockade \cite{Fried2020}.

To study the effect of electric breakdown on the encapsulation layer, we have imaged devices using atomic force microscopy both before and after electric breakdown and found that the  100 nm thick silica-like top layer is not removed by the electric breakdown of the graphene underneath, despite the high temperatures reached during breakdown -- in excess of 1000K \cite{Posa2017, Evangeli2021}. Figure \ref{afm}  shows height profiles on the constrictions. After electric breakdown a slight depression in HSQ height is present in the centre of the constriction, where the temperature is expected to be the highest due to the highest current density, however with a height decrease of 5 nm against the layer thickness of more than 100 nm, the device is still fully encapsulated after the constriction breakdown (figure \ref{afm}). The depression is likely the result of melting or densification of the encapsulating layer. 

%%%%%%%%%%%%%%%%%%%%%%%%
%%%%%%%%%%%%%%%%%%%%%%%%%

We observe two types of conductance switching in the encapsulated devices, both driven by bias voltage ramping. The first one is repeated switching between two conductance states, both of which are in the tunneling regime. The second one is consistent with that observed in \cite{Posa2017, Sarwat2017} -- conductance changes of several orders of magnitude from the tunnelling regime to bulk conductance, which can only be reversed by repeating the electric breakdown process. The large conductance changes have been attributed to either a phase transition,  forming a conductive Si crystal in the substrate \cite{Yao2010, He2012, Posa2017} or geometric restructuring of graphene, with the formation of carbon chains and carbon nanotubes closing the gap \cite{Sarwat2017}. Both effects are expected to occur due to the large field strength in the electroburnt gaps -- in the order of $10^9$V/m. 

An example of the first type of switching is shown in Figure \ref{switching}(a), with a device switching between two similar conductive states (shown in green and blue lines) more than 30 times during bias voltage sweeps from -0.3V to +0.3V. The device switches from the low (green) to the high (blue) conductive state when the bias voltage reaches approximately 0.17V, while the reverse switching is only activated at bias voltages above 0.24V. Following the analysis in \cite{Posa2021}, we have fitted both conductance states at voltage ranges before switching is activated to the Simmons model (see figure \ref{switching}b) and find that the while both states have gap widths of about 2nm, the high state has a gap width approximately 3\AA $ $ lower, while the contact area is 20 times smaller than in the low state. We note that this analysis relies on a continuum model for tunneling, and is not fully applicable to the atomic scale. However, the results are indicative for reproducible atomic scale reaarangements in the nanogap. 

In addition to atomic scale fluctuations, we observe conductance changes of several orders of magnitude. The observed conductance changes are only reversible by subsequent electric breakdown. We explore the nature of this ``breakdown reversal'' by studying the dependence of the conductance of the device after the ``reversal'' on the bias voltage. Prior to electroburning a device conducts through bulk graphene and this function exhibits a dip in conductance at zero bias \cite{Somphonsane2017}, due to weak localisation \cite{Mccann2006, Falko2007} and electron-electron interactions \cite{Kozikov2010}, which is dramatic at low temperatures, and is still observable in our sample at liquid nitrogen temperature (Figure \ref{switching}c).

The zero-bias conductance dip can be used as indication of conductance being mediated by bulk graphene. Figure \ref{electroburning}d shows the typical conductance-voltage trace of a device before the breakdown of the constriction (top) and two types of conductance-voltage traces of devices after breakdown reversal. The typical conductance of a ``reversed'' device observed is $10^{-3}$ of the original conductance of the graphene constriction and the conductance trace exhibits no dip.

In rare cases, we have observed the device switching to a state with even higher conductance: $10^{-2}$ of that of the original graphene constriction, with a conductance-voltage trace demonstrating the same dip at zero bias voltage as bulk graphene before electroburning. Thus, breakdown reversal usually destroys bulk graphene structure. We have fitted the IV trace of a device, that showed bulk graphene behaviour after the ``breakdown reversal'', just before the change in conductance (Figure \ref{switching}c) and found the gap width to be 2 nm -- typical of these devices.

\begin{figure}
\includegraphics[width=0.5\textwidth]{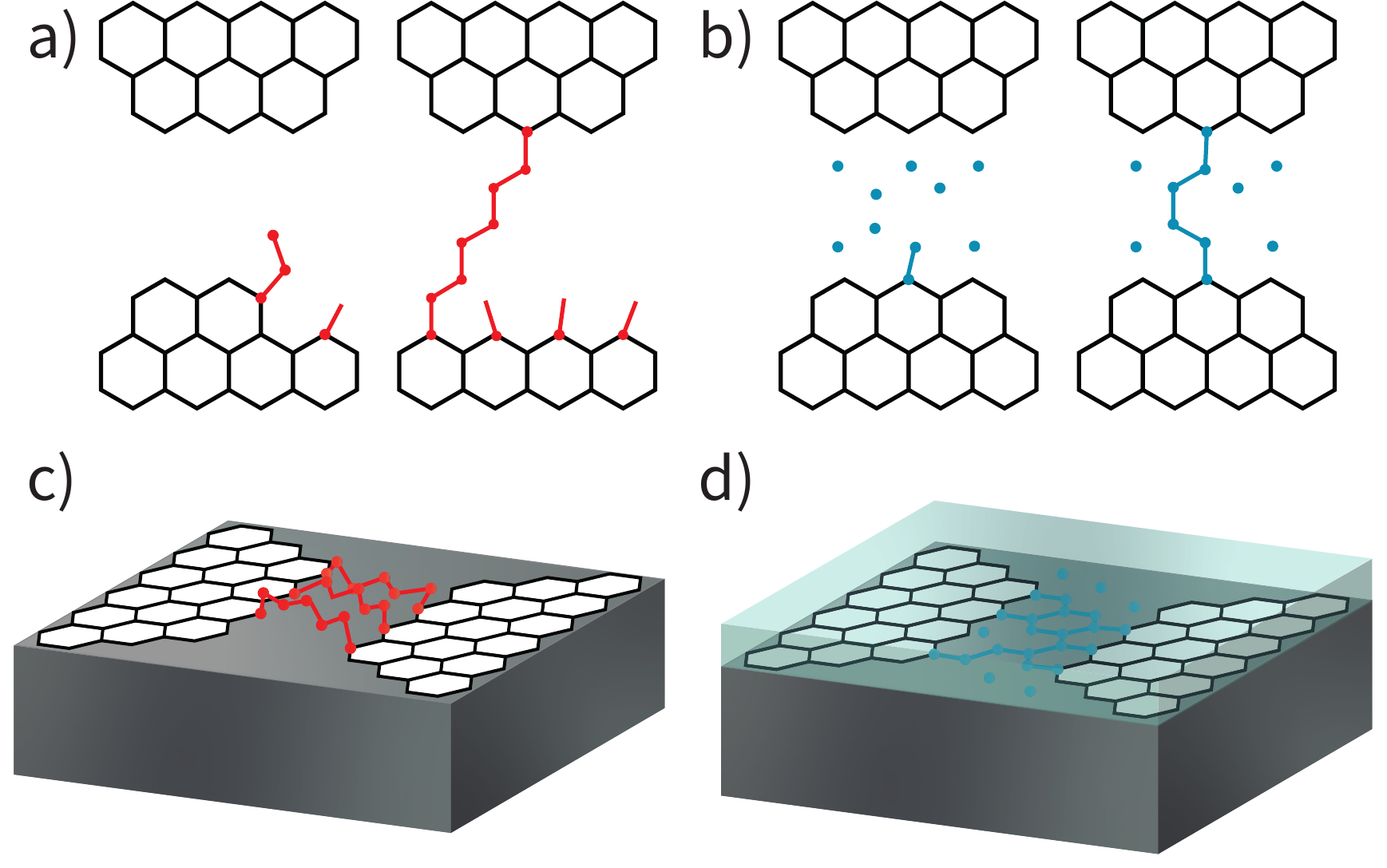}\caption{\label{3dpics} a) The proposed mechanism of conductance switching in a non-encapsulated device. The final row of carbon atoms ``unzips'' under an electric field and can form a chain closing the gap. b) A proposed mechanism of conductance switching in an encapsulated device. As carbon atoms remain in the system after electric breakdown, carbon-carbon bonds can form with atoms in the gap. c) Multiple carbon chains formed in a non-encapsulated gap. As the dynamics is three-dimensional, a disordered structure is likely to form. d) Multiple carbon chains formed in an encapsulated gap. Under two-dimensional dynamics the formation of ordered graphene from adjacent carbon chains is more likely than in the 3D dynamics.}
\end{figure}

We believe that the first type of conductance switching we observed, repeated switching between two states in the tunnelling regime is best described by graphene restructuring. From the Simmons fits of both conductance state we find that the conductance change is due to atomic-scale fluctuations, where the increase of conductance is associated with both a decrease of the fitted gap width and contact area. As the two states are highly reproducible (at the atomic scale), repeated reforming of graphene in the gap, for instance, formation of a protuberance (see Figure \ref{3dpics}a,b) is more likely than the formation and destruction of a silicon crystal identical on the atomic scale.

The large-scale conductance changes, however, can be explained by either graphene restructuring, with the forming carbon chains closing the gap, or the formation of a silicon crystal. Due to the size of the nanogap and the encapsulation, this question cannot easily be solved by microscopy.  It has been demonstrated that the presence of the zero-bias conductance dip in bulk exfoliated graphene depends on the deposition method \cite{Morozov2006}, and its disappearance in some samples is attributed to corrugations in graphene due to imperfect transfer. Therefore, the absence of the zero-bias conductance dip of the devices after breakdown reversal can be attributed to either carbon chains reforming in the gap in a disordered way, or conductance being mediated by a silicon crystal. Similarly, the ``reversed'' devices that demonstrated the zero-bias conductance dip can either have carbon bonds that reformed in an ordered way, forming bulk-like crystaline graphene, (since the conductance of a graphene nanoribbon is proportional to the width down to nanometre-sized widths \cite{Qi2014}, the width of the graphene flake closing the gap is a hundredth of the initial constriction width -- approximately 5 carbon chains that are closely packed together to form sheet graphene), or a very conductive silicon crystal, so that the conductance dip observed is from the quantum effects in the wider parts of the constriction.

Finally, we note that due to the encapsulation, a new mechanism of graphene restructuring may be present in our devices, that does not exist in ``open'' graphene nanogaps. As graphene breakdown occurs underneath a silica layer, carbon atoms cannot escape the system (with the formation of CO$_2$, for instance), and remain in the silica. Therefore, conductance switching may originate from the formation of new bond with the existing carbon sites in the encapsulating layer (Figure \ref{3dpics}b,d) without large-scale movement of the carbon atoms, unlike the mechanism described in \cite{Sarwat2017} (Figure \ref{3dpics}a,c). Additionally, it is plausible that the restriction of dynamics to two dimensions favours the reformation of ordered graphene (Figure \ref{3dpics}c,d).

%%%%%%%%%%%%%%%%%%
%%%%%%%%%%%%%%%%%

To conclude, we investigated the feasibility of fabricating graphene constrictions for further feedback-controlled electric breakdown using HSQ resist and find that the presence of the passivation layer of silica-like material remaining after exposure and development of the resist, does not affect the breakdown process and both nanogaps and quantum dots can be formed. In turn, the Joule heat dissipated during feedback-controlled breakdown does not affect the encapsulating silica layer in a detrimental way.

As a result, we are able to fabricate nanometre-scale graphene nanogaps and quantum dots with addition energies large enough to show room-temperature Coulomb blockade, using a common negative resist. This approach offers technological benefits as the devices being covered by a layer of  SiO$_2$ of thickness which can be varied through initial spin coating. These include protection against environmental conditions and the possibility of placing additional electrodes over the silica layer, allowing the placement of gate electrodes much closer to the quantum dot than a back or a side-gate. Side gate electrodes can't be placed closer than the original width of the constriction \cite{Puczkarski2015}, approximately 100 nm, while the HSQ layer can be down to 10nm thick. Close placement of gate electrodes reduces the voltage needed to gate a single-electron transistor compared to using a back gate. Additionally, in thermoelectric devices, reducing the distance between the heater and the device (made possible by the top heater placement) solves the problem of the large dispersed heater powers\cite{Gehring2019}.

We have explored conductance switching in HSQ-encapsulated devices and found that they demonstrate reproducible bias-voltage driven conductance changes in the tunnelling regime due to atomic-scale fluctuations in the nanogap, as well as previously described switches from tunnelling to bulk conductance. We attribute the fluctuations in the encapsulated nanogap to the geometric restructuring of graphene under electric field. We believe that studying the physical mechanism behind the conductance switching in nanodevices can provide insights on both the scaling limits of nanoelectronics and nanoscale memory elements.

%\bibliography{refs}% Produces the bibliography via BibTeX.
%

\end{document}